\documentstyle[12pt]{article}
\newcommand {\cD}{{\cal D}}

\def\a{\alpha}
\def\b{\beta}
\def\c{\chi}
\def\d{\delta}

\def\k{\kappa}
\def\l{\lambda}
\def\m{\mu}
\def\n{\nu}

\def\r{\rho}

\def\t{\tau}

\def\x{\xi}

\def\L{\Lambda}

\newcommand{\be}{\begin{equation}}
\newcommand{\ee}{\end{equation}}
\newcommand{\bea}{\begin{eqnarray}}
\newcommand{\eea}{\end{eqnarray}}
\newcommand{\non}{\nonumber}
\title{${\cal N}$=4 Super Yang-Mills Low-Energy Effective Action at
Three and Four Loops}
             \author{I.L. Buchbinder$^{a,b}$,  A.Yu.
Petrov$^{a}$\\ $^{a}$ \small\it Department of Theoretical Physics\\
\small\it Tomsk
State Pedagogical University\\ \small\it Tomsk, 634041, Russia\\
$^{b}$
\small\it Instituto de F\'{i}sica, Universidade de S\~{a}o Paulo\\
\small\it P.O.Box 66318,
05315-970, S\~{a}o Paulo, Brasil \normalsize\rm}
\date{}
\begin{document}
\immediate\write16{<WARNING: FEYNMAN macros work only with
emTeX-dvivers (dviscr.exe, dvihplj.exe, dvidot.exe, etc.) >}
\newdimen\Lengthunit
\newcount\Nhalfperiods
\Lengthunit = 1.5cm
\Nhalfperiods = 9
\catcode`\*=11
\newdimen\L*   \newdimen\d*   \newdimen\d**
\newdimen\dm*  \newdimen\dd*  \newdimen\dt*
\newdimen\a*   \newdimen\b*   \newdimen\c*
\newdimen\a**  \newdimen\b**
\newdimen\xL*  \newdimen\yL*
\newcount\k*   \newcount\l*   \newcount\m*
\newcount\n*   \newcount\dn*  \newcount\r*
\newcount\N*   \newcount\*one \newcount\*two  \*one=1 \*two=2
\newcount\*ths \*ths=1000
\def\GRAPH(hsize=#1)#2{\hbox to #1\Lengthunit{#2\hss}}
\def\Linewidth#1{\special{em:linewidth #1}}
\Linewidth{.4pt}
\def\sm*{\special{em:moveto}}
\def\sl*{\special{em:lineto}}
\newbox\spm*   \newbox\spl*
\setbox\spm*\hbox{\sm*}
\setbox\spl*\hbox{\sl*}
\def\mov#1(#2,#3)#4{\rlap{\L*=#1\Lengthunit\kern#2\L*\raise#3\L*\hbox{#4}}}
\def\smov#1(#2,#3)#4{\rlap{\L*=#1\Lengthunit
\xL*=\xscale\L*\yL*=\yscale\L*\kern#2\xL*\raise#3\yL*\hbox{#4}}}
\def\mov*(#1,#2)#3{\rlap{\kern#1\raise#2\hbox{#3}}}
\def\lin#1(#2,#3){\rlap{\sm*\mov#1(#2,#3){\sl*}}}
\def\arr*(#1,#2,#3){\mov*(#1\dd*,#1\dt*){%
\sm*\mov*(#2\dd*,#2\dt*){\mov*(#3\dt*,-#3\dd*){\sl*}}%
\sm*\mov*(#2\dd*,#2\dt*){\mov*(-#3\dt*,#3\dd*){\sl*}}}}
\def\arrow#1(#2,#3){\rlap{\lin#1(#2,#3)\mov#1(#2,#3){%
\d**=-.012\Lengthunit\dd*=#2\d**\dt*=#3\d**%
\arr*(1,10,4)\arr*(3,8,4)\arr*(4.8,4.2,3)}}}
\def\arrlin#1(#2,#3){\rlap{\L*=#1\Lengthunit\L*=.5\L*%
\lin#1(#2,#3)\mov*(#2\L*,#3\L*){\arrow.1(#2,#3)}}}
\def\dasharrow#1(#2,#3){\rlap{%
{\Lengthunit=0.9\Lengthunit\dashlin#1(#2,#3)\mov#1(#2,#3){\sm*}}%
\mov#1(#2,#3){\sl*\d**=-.012\Lengthunit\dd*=#2\d**\dt*=#3\d**%
\arr*(1,10,4)\arr*(3,8,4)\arr*(4.8,4.2,3)}}}
\def\clap#1{\hbox to 0pt{\hss #1\hss}}
\def\ind(#1,#2)#3{\rlap{%
\d*=.1\Lengthunit\kern#1\d*\raise#2\d*\hbox{\lower2pt\clap{$#3$}}}}
\def\sh*(#1,#2)#3{\rlap{%
\dm*=\the\n*\d**\xL*=\xscale\dm*\yL*=\yscale\dm*
\kern#1\xL*\raise#2\yL*\hbox{#3}}}
\def\calcnum*#1(#2,#3){\a*=1000sp\b*=1000sp\a*=#2\a*\b*=#3\b*%
\ifdim\a*<0pt\a*-\a*\fi\ifdim\b*<0pt\b*-\b*\fi%
\ifdim\a*>\b*\c*=.96\a*\advance\c*.4\b*%
\else\c*=.96\b*\advance\c*.4\a*\fi%
\k*\a*\multiply\k*\k*\l*\b*\multiply\l*\l*%
\m*\k*\advance\m*\l*\n*\c*\r*\n*\multiply\n*\n*%
\dn*\m*\advance\dn*-\n*\divide\dn*2\divide\dn*\r*%
\advance\r*\dn*%
\c*=\the\Nhalfperiods5sp\c*=#1\c*\ifdim\c*<0pt\c*-\c*\fi%
\multiply\c*\r*\N*\c*\divide\N*10000}
\def\dashlin#1(#2,#3){\rlap{\calcnum*#1(#2,#3)%
\d**=#1\Lengthunit\ifdim\d**<0pt\d**-\d**\fi%
\divide\N*2\multiply\N*2\advance\N*1%
\divide\d**\N*\sm*\n*\*one\sh*(#2,#3){\sl*}%
\loop\advance\n*\*one\sh*(#2,#3){\sm*}\advance\n*\*one\sh*(#2,#3){\sl*}%
\ifnum\n*<\N*\repeat}}
\def\dashdotlin#1(#2,#3){\rlap{\calcnum*#1(#2,#3)%
\d**=#1\Lengthunit\ifdim\d**<0pt\d**-\d**\fi%
\divide\N*2\multiply\N*2\advance\N*1\multiply\N*2%
\divide\d**\N*\sm*\n*\*two\sh*(#2,#3){\sl*}\loop%
\advance\n*\*one\sh*(#2,#3){\kern-1.48pt\lower.5pt\hbox{\rm.}}%
\advance\n*\*one\sh*(#2,#3){\sm*}%
\advance\n*\*two\sh*(#2,#3){\sl*}\ifnum\n*<\N*\repeat}}
\def\shl*(#1,#2)#3{\kern#1#3\lower#2#3\hbox{\unhcopy\spl*}}
\def\trianglin#1(#2,#3){\rlap{\toks0={#2}\toks1={#3}\calcnum*#1(#2,#3)%
\dd*=.57\Lengthunit\dd*=#1\dd*\divide\dd*\N*%
\d**=#1\Lengthunit\ifdim\d**<0pt\d**-\d**\fi%
\multiply\N*2\divide\d**\N*\advance\N*-1\sm*\n*\*one\loop%
\shl**{\dd*}\dd*-\dd*\advance\n*2%
\ifnum\n*<\N*\repeat\n*\N*\advance\n*1\shl**{0pt}}}
\def\wavelin#1(#2,#3){\rlap{\toks0={#2}\toks1={#3}\calcnum*#1(#2,#3)%
\dd*=.23\Lengthunit\dd*=#1\dd*\divide\dd*\N*%
\d**=#1\Lengthunit\ifdim\d**<0pt\d**-\d**\fi%
\multiply\N*4\divide\d**\N*\sm*\n*\*one\loop%
\shl**{\dd*}\dt*=1.3\dd*\advance\n*1%
\shl**{\dt*}\advance\n*\*one%
\shl**{\dd*}\advance\n*\*two%
\dd*-\dd*\ifnum\n*<\N*\repeat\n*\N*\shl**{0pt}}}
\def\w*lin(#1,#2){\rlap{\toks0={#1}\toks1={#2}\d**=\Lengthunit\dd*=-.12\d**%
\N*8\divide\d**\N*\sm*\n*\*one\loop%
\shl**{\dd*}\dt*=1.3\dd*\advance\n*\*one%
\shl**{\dt*}\advance\n*\*one%
\shl**{\dd*}\advance\n*\*one%
\shl**{0pt}\dd*-\dd*\advance\n*1\ifnum\n*<\N*\repeat}}
\def\l*arc(#1,#2)[#3][#4]{\rlap{\toks0={#1}\toks1={#2}\d**=\Lengthunit%
\dd*=#3.037\d**\dd*=#4\dd*\dt*=#3.049\d**\dt*=#4\dt*\ifdim\d**>16mm%
\d**=.25\d**\n*\*one\shl**{-\dd*}\n*\*two\shl**{-\dt*}\n*3\relax%
\shl**{-\dd*}\n*4\relax\shl**{0pt}\else\ifdim\d**>5mm%
\d**=.5\d**\n*\*one\shl**{-\dt*}\n*\*two\shl**{0pt}%
\else\n*\*one\shl**{0pt}\fi\fi}}
\def\d*arc(#1,#2)[#3][#4]{\rlap{\toks0={#1}\toks1={#2}\d**=\Lengthunit%
\dd*=#3.037\d**\dd*=#4\dd*\d**=.25\d**\sm*\n*\*one\shl**{-\dd*}%
\n*3\relax\sh*(#1,#2){\xL*=\xscale\dd*\yL*=\yscale\dd*
\kern#2\xL*\lower#1\yL*\hbox{\sm*}}%
\n*4\relax\shl**{0pt}}}
\def\arc#1[#2][#3]{\rlap{\Lengthunit=#1\Lengthunit%
\sm*\l*arc(#2.1914,#3.0381)[#2][#3]%
\smov(#2.1914,#3.0381){\l*arc(#2.1622,#3.1084)[#2][#3]}%
\smov(#2.3536,#3.1465){\l*arc(#2.1084,#3.1622)[#2][#3]}%
\smov(#2.4619,#3.3086){\l*arc(#2.0381,#3.1914)[#2][#3]}}}
\def\dasharc#1[#2][#3]{\rlap{\Lengthunit=#1\Lengthunit%
\d*arc(#2.1914,#3.0381)[#2][#3]%
\smov(#2.1914,#3.0381){\d*arc(#2.1622,#3.1084)[#2][#3]}%
\smov(#2.3536,#3.1465){\d*arc(#2.1084,#3.1622)[#2][#3]}%
\smov(#2.4619,#3.3086){\d*arc(#2.0381,#3.1914)[#2][#3]}}}
\def\wavearc#1[#2][#3]{\rlap{\Lengthunit=#1\Lengthunit%
\w*lin(#2.1914,#3.0381)%
\smov(#2.1914,#3.0381){\w*lin(#2.1622,#3.1084)}%
\smov(#2.3536,#3.1465){\w*lin(#2.1084,#3.1622)}%
\smov(#2.4619,#3.3086){\w*lin(#2.0381,#3.1914)}}}
\def\shl**#1{\c*=\the\n*\d**\d*=#1%
\a*=\the\toks0\c*\b*=\the\toks1\d*\advance\a*-\b*%
\b*=\the\toks1\c*\d*=\the\toks0\d*\advance\b*\d*%
\a*=\xscale\a*\b*=\yscale\b*%
\raise\b*\rlap{\kern\a*\unhcopy\spl*}}
\def\wlin*#1(#2,#3)[#4]{\rlap{\toks0={#2}\toks1={#3}%
\c*=#1\l*\c*\c*=.01\Lengthunit\m*\c*\divide\l*\m*%
\c*=\the\Nhalfperiods5sp\multiply\c*\l*\N*\c*\divide\N*\*ths%
\divide\N*2\multiply\N*2\advance\N*1%
\dd*=.002\Lengthunit\dd*=#4\dd*\multiply\dd*\l*\divide\dd*\N*%
\d**=#1\multiply\N*4\divide\d**\N*\sm*\n*\*one\loop%
\shl**{\dd*}\dt*=1.3\dd*\advance\n*\*one%
\shl**{\dt*}\advance\n*\*one%
\shl**{\dd*}\advance\n*\*two%
\dd*-\dd*\ifnum\n*<\N*\repeat\n*\N*\shl**{0pt}}}
\def\wavebox#1{\setbox0\hbox{#1}%
\a*=\wd0\advance\a*14pt\b*=\ht0\advance\b*\dp0\advance\b*14pt%
\hbox{\kern9pt%
\mov*(0pt,\ht0){\mov*(-7pt,7pt){\wlin*\a*(1,0)[+]\wlin*\b*(0,-1)[-]}}%
\mov*(\wd0,-\dp0){\mov*(7pt,-7pt){\wlin*\a*(-1,0)[+]\wlin*\b*(0,1)[-]}}%
\box0\kern9pt}}
\def\rectangle#1(#2,#3){%
\lin#1(#2,0)\lin#1(0,#3)\mov#1(0,#3){\lin#1(#2,0)}\mov#1(#2,0){\lin#1(0,#3)}}
\def\dashrectangle#1(#2,#3){\dashlin#1(#2,0)\dashlin#1(0,#3)%
\mov#1(0,#3){\dashlin#1(#2,0)}\mov#1(#2,0){\dashlin#1(0,#3)}}
\def\waverectangle#1(#2,#3){\L*=#1\Lengthunit\a*=#2\L*\b*=#3\L*%
\ifdim\a*<0pt\a*-\a*\def\x*{-1}\else\def\x*{1}\fi%
\ifdim\b*<0pt\b*-\b*\def\y*{-1}\else\def\y*{1}\fi%
\wlin*\a*(\x*,0)[-]\wlin*\b*(0,\y*)[+]%
\mov#1(0,#3){\wlin*\a*(\x*,0)[+]}\mov#1(#2,0){\wlin*\b*(0,\y*)[-]}}
\def\calcparab*{%
\ifnum\n*>\m*\k*\N*\advance\k*-\n*\else\k*\n*\fi%
\a*=\the\k* sp\a*=10\a*\b*\dm*\advance\b*-\a*\k*\b*%
\a*=\the\*ths\b*\divide\a*\l*\multiply\a*\k*%
\divide\a*\l*\k*\*ths\r*\a*\advance\k*-\r*%
\dt*=\the\k*\L*}
\def\arcto#1(#2,#3)[#4]{\rlap{\toks0={#2}\toks1={#3}\calcnum*#1(#2,#3)%
\dm*=135sp\dm*=#1\dm*\d**=#1\Lengthunit\ifdim\dm*<0pt\dm*-\dm*\fi%
\multiply\dm*\r*\a*=.3\dm*\a*=#4\a*\ifdim\a*<0pt\a*-\a*\fi%
\advance\dm*\a*\N*\dm*\divide\N*10000%
\divide\N*2\multiply\N*2\advance\N*1%
\L*=-.25\d**\L*=#4\L*\divide\d**\N*\divide\L*\*ths%
\m*\N*\divide\m*2\dm*=\the\m*5sp\l*\dm*%
\sm*\n*\*one\loop\calcparab*\shl**{-\dt*}%
\advance\n*1\ifnum\n*<\N*\repeat}}
\def\arrarcto#1(#2,#3)[#4]{\L*=#1\Lengthunit\L*=.54\L*%
\arcto#1(#2,#3)[#4]\mov*(#2\L*,#3\L*){\d*=.457\L*\d*=#4\d*\d**-\d*%
\mov*(#3\d**,#2\d*){\arrow.02(#2,#3)}}}
\def\dasharcto#1(#2,#3)[#4]{\rlap{\toks0={#2}\toks1={#3}\calcnum*#1(#2,#3)%
\dm*=\the\N*5sp\a*=.3\dm*\a*=#4\a*\ifdim\a*<0pt\a*-\a*\fi%
\advance\dm*\a*\N*\dm*%
\divide\N*20\multiply\N*2\advance\N*1\d**=#1\Lengthunit%
\L*=-.25\d**\L*=#4\L*\divide\d**\N*\divide\L*\*ths%
\m*\N*\divide\m*2\dm*=\the\m*5sp\l*\dm*%
\sm*\n*\*one\loop%
\calcparab*\shl**{-\dt*}\advance\n*1%
\ifnum\n*>\N*\else\calcparab*%
\sh*(#2,#3){\kern#3\dt*\lower#2\dt*\hbox{\sm*}}\fi%
\advance\n*1\ifnum\n*<\N*\repeat}}
\def\*shl*#1{%
\c*=\the\n*\d**\advance\c*#1\a**\d*\dt*\advance\d*#1\b**%
\a*=\the\toks0\c*\b*=\the\toks1\d*\advance\a*-\b*%
\b*=\the\toks1\c*\d*=\the\toks0\d*\advance\b*\d*%
\raise\b*\rlap{\kern\a*\unhcopy\spl*}}
\def\calcnormal*#1{%
\b**=10000sp\a**\b**\k*\n*\advance\k*-\m*%
\multiply\a**\k*\divide\a**\m*\a**=#1\a**\ifdim\a**<0pt\a**-\a**\fi%
\ifdim\a**>\b**\d*=.96\a**\advance\d*.4\b**%
\else\d*=.96\b**\advance\d*.4\a**\fi%
\d*=.01\d*\r*\d*\divide\a**\r*\divide\b**\r*%
\ifnum\k*<0\a**-\a**\fi\d*=#1\d*\ifdim\d*<0pt\b**-\b**\fi%
\k*\a**\a**=\the\k*\dd*\k*\b**\b**=\the\k*\dd*}
\def\wavearcto#1(#2,#3)[#4]{\rlap{\toks0={#2}\toks1={#3}\calcnum*#1(#2,#3)%
\c*=\the\N*5sp\a*=.4\c*\a*=#4\a*\ifdim\a*<0pt\a*-\a*\fi%
\advance\c*\a*\N*\c*\divide\N*20\multiply\N*2\advance\N*-1\multiply\N*4%
\d**=#1\Lengthunit\dd*=.012\d**\ifdim\d**<0pt\d**-\d**\fi\L*=.25\d**%
\divide\d**\N*\divide\dd*\N*\L*=#4\L*\divide\L*\*ths%
\m*\N*\divide\m*2\dm*=\the\m*0sp\l*\dm*%
\sm*\n*\*one\loop\calcnormal*{#4}\calcparab*%
\*shl*{1}\advance\n*\*one\calcparab*%
\*shl*{1.3}\advance\n*\*one\calcparab*%
\*shl*{1}\advance\n*2%
\dd*-\dd*\ifnum\n*<\N*\repeat\n*\N*\shl**{0pt}}}
\def\triangarcto#1(#2,#3)[#4]{\rlap{\toks0={#2}\toks1={#3}\calcnum*#1(#2,#3)%
\c*=\the\N*5sp\a*=.4\c*\a*=#4\a*\ifdim\a*<0pt\a*-\a*\fi%
\advance\c*\a*\N*\c*\divide\N*20\multiply\N*2\advance\N*-1\multiply\N*2%
\d**=#1\Lengthunit\dd*=.012\d**\ifdim\d**<0pt\d**-\d**\fi\L*=.25\d**%
\divide\d**\N*\divide\dd*\N*\L*=#4\L*\divide\L*\*ths%
\m*\N*\divide\m*2\dm*=\the\m*0sp\l*\dm*%
\sm*\n*\*one\loop\calcnormal*{#4}\calcparab*%
\*shl*{1}\advance\n*2%
\dd*-\dd*\ifnum\n*<\N*\repeat\n*\N*\shl**{0pt}}}
\def\hr*#1{\clap{\xL*=\xscale\Lengthunit\vrule width#1\xL*
height.1pt}}
\def\shade#1[#2]{\rlap{\Lengthunit=#1\Lengthunit%
\smov(0,#2.05){\hr*{.994}}\smov(0,#2.1){\hr*{.980}}%
\smov(0,#2.15){\hr*{.953}}\smov(0,#2.2){\hr*{.916}}%
\smov(0,#2.25){\hr*{.867}}\smov(0,#2.3){\hr*{.798}}%
\smov(0,#2.35){\hr*{.715}}\smov(0,#2.4){\hr*{.603}}%
\smov(0,#2.45){\hr*{.435}}}}
\def\dshade#1[#2]{\rlap{%
\Lengthunit=#1\Lengthunit\if#2-\def\t*{+}\else\def\t*{-}\fi%
\smov(0,\t*.025){%
\smov(0,#2.05){\hr*{.995}}\smov(0,#2.1){\hr*{.988}}%
\smov(0,#2.15){\hr*{.969}}\smov(0,#2.2){\hr*{.937}}%
\smov(0,#2.25){\hr*{.893}}\smov(0,#2.3){\hr*{.836}}%
\smov(0,#2.35){\hr*{.760}}\smov(0,#2.4){\hr*{.662}}%
\smov(0,#2.45){\hr*{.531}}\smov(0,#2.5){\hr*{.320}}}}}
\def\vdot{\rlap{\kern-1.9pt\lower1.8pt\hbox{$\scriptstyle\bullet$}}}
\def\vtimes{\rlap{\kern-3pt\lower1.8pt\hbox{$\scriptstyle\times$}}}
\def\vDot{\rlap{\kern-2.3pt\lower2.7pt\hbox{$\bullet$}}}
\def\vTimes{\rlap{\kern-3.6pt\lower2.4pt\hbox{$\times$}}}
\catcode`\*=12
\newcount\CatcodeOfAtSign
\CatcodeOfAtSign=\the\catcode`\@
\catcode`\@=11
\newcount\n@ast
\def\n@ast@#1{\n@ast0\relax\get@ast@#1\end}
\def\get@ast@#1{\ifx#1\end\let\next\relax\else%
\ifx#1*\advance\n@ast1\fi\let\next\get@ast@\fi\next}
\newif\if@up \newif\if@dwn
\def\up@down@#1{\@upfalse\@dwnfalse%
\if#1u\@uptrue\fi\if#1U\@uptrue\fi\if#1+\@uptrue\fi%
\if#1d\@dwntrue\fi\if#1D\@dwntrue\fi\if#1-\@dwntrue\fi}
\def\halfcirc#1(#2)[#3]{{\Lengthunit=#2\Lengthunit\up@down@{#3}%
\if@up\smov(0,.5){\arc[-][-]\arc[+][-]}\fi%
\if@dwn\smov(0,-.5){\arc[-][+]\arc[+][+]}\fi%
\def\lft{\smov(0,.5){\arc[-][-]}\smov(0,-.5){\arc[-][+]}}%
\def\rght{\smov(0,.5){\arc[+][-]}\smov(0,-.5){\arc[+][+]}}%
\if#3l\lft\fi\if#3L\lft\fi\if#3r\rght\fi\if#3R\rght\fi%
\n@ast@{#1}%
\ifnum\n@ast>0\if@up\shade[+]\fi\if@dwn\shade[-]\fi\fi%
\ifnum\n@ast>1\if@up\dshade[+]\fi\if@dwn\dshade[-]\fi\fi}}
\def\halfdashcirc(#1)[#2]{{\Lengthunit=#1\Lengthunit\up@down@{#2}%
\if@up\smov(0,.5){\dasharc[-][-]\dasharc[+][-]}\fi%
\if@dwn\smov(0,-.5){\dasharc[-][+]\dasharc[+][+]}\fi%
\def\lft{\smov(0,.5){\dasharc[-][-]}\smov(0,-.5){\dasharc[-][+]}}%
\def\rght{\smov(0,.5){\dasharc[+][-]}\smov(0,-.5){\dasharc[+][+]}}%
\if#2l\lft\fi\if#2L\lft\fi\if#2r\rght\fi\if#2R\rght\fi}}
\def\halfwavecirc(#1)[#2]{{\Lengthunit=#1\Lengthunit\up@down@{#2}%
\if@up\smov(0,.5){\wavearc[-][-]\wavearc[+][-]}\fi%
\if@dwn\smov(0,-.5){\wavearc[-][+]\wavearc[+][+]}\fi%
\def\lft{\smov(0,.5){\wavearc[-][-]}\smov(0,-.5){\wavearc[-][+]}}%
\def\rght{\smov(0,.5){\wavearc[+][-]}\smov(0,-.5){\wavearc[+][+]}}%
\if#2l\lft\fi\if#2L\lft\fi\if#2r\rght\fi\if#2R\rght\fi}}
\def\Circle#1(#2){\halfcirc#1(#2)[u]\halfcirc#1(#2)[d]\n@ast@{#1}%
\ifnum\n@ast>0\clap{%
\dimen0=\xscale\Lengthunit\vrule width#2\dimen0 height.1pt}\fi}
\def\wavecirc(#1){\halfwavecirc(#1)[u]\halfwavecirc(#1)[d]}
\def\dashcirc(#1){\halfdashcirc(#1)[u]\halfdashcirc(#1)[d]}
%
\def\xscale{1}
\def\yscale{1}
\def\Ellipse#1(#2)[#3,#4]{\def\xscale{#3}\def\yscale{#4}%
\Circle#1(#2)\def\xscale{1}\def\yscale{1}}
\def\dashEllipse(#1)[#2,#3]{\def\xscale{#2}\def\yscale{#3}%
\dashcirc(#1)\def\xscale{1}\def\yscale{1}}
\def\waveEllipse(#1)[#2,#3]{\def\xscale{#2}\def\yscale{#3}%
\wavecirc(#1)\def\xscale{1}\def\yscale{1}}
\def\halfEllipse#1(#2)[#3][#4,#5]{\def\xscale{#4}\def\yscale{#5}%
\halfcirc#1(#2)[#3]\def\xscale{1}\def\yscale{1}}
\def\halfdashEllipse(#1)[#2][#3,#4]{\def\xscale{#3}\def\yscale{#4}%
\halfdashcirc(#1)[#2]\def\xscale{1}\def\yscale{1}}
\def\halfwaveEllipse(#1)[#2][#3,#4]{\def\xscale{#3}\def\yscale{#4}%
\halfwavecirc(#1)[#2]\def\xscale{1}\def\yscale{1}}
\catcode`\@=\the\CatcodeOfAtSign
\maketitle
\begin{abstract}
We investigate the low-energy effective action in N=4 super
Yang-Mills
theory with gauge group $SU(n)$ spontaneously broken down to its
maximal
torus. Using harmonic superspace technique we prove an absence of any
three- and four-loop corrections to non-holomorphic effective
potential
depending on $N=2$ superfield strengths. A mechanism responsible
for vanishing arbitrary loop corrections to low-energy effective
action is
discussed.
\end{abstract}

Supersymmetry imposes the significant restrictions on a structure of
effective action in field models. It is naturally to expect that the
most
strong restrictions have to arise in maximally extended rigid
supersymmetric model, that is in $N=4$ super-Yang-Mills theory.

Recently Dine and Seiberg found that a dependence of $N=4$
supersymmetric Yang-Mills low-energy effective action on $N=2$
superfield strengths ${\cal W}$ and $\bar{\cal W}$ is exactly fixed
only by general properties of the quantum theory under consideration
like finiteness and scale independence [1].  According to ref [1] the
leading low-energy contributions to effective action in $N=4$ SYM
with
gauge group $SU(2)$ spontaneously broken down to $U(1)$ are
described by
non-holomorphic effective potential ${\cal H}({\cal W},\bar{\cal W})$
of the form
\bea
{\cal H}({\cal W},\bar{\cal W}) =
c\log\big(\frac{{\cal W}^2}{\Lambda^2}\big)
\log\big(\frac{\bar{\cal W}^2}{\Lambda^2}\big)
\eea
Here $\Lambda$ is some scale. The effective
potential (1) possesses by two remarkable properties. First, the
corresponding effective action
\bea
\int d^{4}xd^{8}\theta {\cal H}({\cal W},\bar{\cal W})
\eea
is scale independent. Second, any
quantum corrections, if they exist at all, are included into a single
constant $c$.

The explicit calculations of the non-holomorphic effective potential
and
finding the coefficient $c$ in one-loop approximation have been
carried out
in refs [2-4]. Extension of the above results for $N=4$ Yang-Mills
theory with gauge group $SU(n)$, $n>2$, spontaneously
broken down to its maximal torus have been developed in refs [5-8].
General structure of low-energy effective action in $N=2,4$
superconformal invariant field models was investigated in ref [9].

In the paper [1] Dine and Seiberg presented the qualitative
arguments based on principle of naturalness [10] (see application of
this
principle to SUSY theories in ref [11]) that effective potential (1)
gets neither
perturbative nor non-perturbative quantum corrections beyond one
loop.
Therefore the expression (1) together with the results of one-loop
calculations of the coefficient $c$ [2-4] determines exact low-energy
effective action in $N=4$ $SU(2)$ Yang-Mills theory. The above
arguments
have also been discussed in refs [8,15]. Another approach leading to the
same conclusion about structure of low-energy effective action was
developed in recent papers [22].

We would like to pay an attention that a mechanism providing an
absence
of higher loop corrections to non-holomorphic effective potential
in $N=4$ Yang-Mills theory is unknown up to now. The firm results
concern only two-loop approximation where the corresponding
corrections
are prohibited by $N=2$ supersymmetry [12] (see also direct two-loop
calculations in refs [13,14]).

An interesting aspect of $N=4$ Yang-Mills theory with gauge group
$SU(n)$, $n>2$, has been recently pointed out in refs [8,15].
The symmetry arguments do not prohibit an appearance of some new
invariant structures, besides logarithmic, in non-holomorphic
effective
potential which are absent at $n=2$. The direct calculations [5-8] do
not confirm such structures in one-loop approximation. However a
question concerning their appearance at higher loops is open.

In this paper we are going to develop a technique for investigating a
structure of the non-holomorphic effective potential at higher
loops, to
find a mechanism providing a cancellation of higher-loop
contributions, and to
clarify situation concerning the non-logarithimic corrections to
low-energy effective action in $N=4$ Yang-Mills theory with gauge
group
$SU(n)$, $n>2$, spontaneously broken to its maximal torus. To be
more precise, we investigate a structure of three- and four-loop
supergraphs and show how $N=4$ supersymmetry provides an efficient
mechanism of supergraph cancellations.

We consider $N=4$ Yang-Mills theory formulated in terms of $N=2$
superfields and get $N=2$ Yang-Mills theory coupled to hypermultiplet
in adjoint representation. The most convenient and simple way to
carry
out quantum calculations in $N=2$ SUSY models is given by harmonic
superspace approach [16-18] which is used in the paper. The various
implementations of this approach to effective action in $N=2$ SUSY
theories are discussed in refs [3,7,12,19,20].

The starting point of our consideration is the classical
action of the $N=4$ super-Yang-Mills theory in $q$-hypermultiplet
realization written in harmonic superspace [16,17]
\begin{eqnarray}
S&=&\frac{1}{g^2} {\rm tr}\,
\int d^{12}z\sum\limits_{n=2}^\infty\frac{(-{\rm i})^n} {n}\int du_1
\cdots du_n\frac{V^{++}(z,u_1)\cdots V^{++}(z,u_n)} {(u^+_1
u^+_2)(u^+_2 u^+_3)\cdots (u^+_n u^+_1)}+\nonumber\\&+&
\int d\zeta^{(-4)} du\breve{q}^{+}(D^{++}+iV^{++})q^{+}.  \label{16}
\end{eqnarray}
The denotions introduced in the
paper \cite{bko} are employed here and further.

The calculations are carried out in framework of $N=2$
background field method \cite{bbko}.
We make background-quantum splitting by the rule
\begin{equation}
V^{++}\rightarrow V^{++}+g\,v^{++} \label{18}
\end{equation}
and construct the corresponding Faddeev-Popov ghost action in the
form \cite{bko}
\begin{equation}
\label{gh}
S_{gh}={\rm tr}\int d\zeta^{(-4)} du
{\bf b}(\nabla^{++})^2{\bf c} -
{\rm i}\,g\, {\rm tr} \,\int du
\,d\zeta^{(-4)}\, \nabla^{++}{\bf b}\;[v^{++}, {\bf c}]\;.
\end{equation}

The background-dependent superpropagators in the theory with action
of $N=2$
gauge multiplet and $N=2$ matter hypermultiplet (\ref{16}) and
action of
ghosts (\ref{gh}) have been obtained in \cite{bko} and look like
\begin{eqnarray}
<v^{++}_\t(1)\,v^{++}_\t(2)>
&=& - \frac{{\rm i}}{{\stackrel{\frown}{\Box}}{}}
{\stackrel{\longrightarrow}{(\cD_1^+)^4}}{}
\biggl\{ \delta^{12}(z_1-z_2) \d^{(-2,2)}(u_1,u_2) \biggr\}
\non \\
<q^+_\t(1)\,\breve{q}^+_\t(2)>
&=& \; \frac{{\rm i}}{{\stackrel{\frown}{\Box}}{}}
{\stackrel{\longrightarrow}{(\cD_1^+)^4}}{}
\left\{ \delta^{12}(z_1-z_2)
{1\over (u^+_1 u^+_2)^3} \right\}
{\stackrel{\longleftarrow}{(\cD_2^+)^4}}
\non \\
<\omega_\t(1)\,\,\omega^{\rm T}_\t(2)>
&=& - \frac{{\rm i}}{{\stackrel{\frown}{\Box}}{}}
{\stackrel{\longrightarrow}{(\cD_1^+)^4}}{}
\left\{ \delta^{12}(z_1-z_2)
{(u^-_1 u^-_2)\over (u^+_1 u^+_2)^3}
\right\}
{\stackrel{\longleftarrow}{(\cD_2^+)^4}}
\non \\
<{\bf c}_\t(1)\,\,{\bf b}_\t(2)>
&=& - \frac{{\rm i}}{{\stackrel{\frown}{\Box}}{}}
{\stackrel{\longrightarrow}{(\cD_1^+)^4}}{}
\left\{ \delta^{12}(z_1-z_2)
{(u^-_1 u^-_2)\over (u^+_1 u^+_2)^3}
\right\}
{\stackrel{\longleftarrow}{(\cD_2^+)^4}}\;.
\label{28}
\end{eqnarray}
with the operator ${\stackrel{\frown}{\Box}}$ of the form \cite{bko}
\begin{eqnarray}
{\stackrel{\frown}{\Box}}{}&=&
{\cal D}^m{\cal D}_m+
\frac{{\rm i}}{2}({\cal D}^{+\alpha}{\cal W}){\cal
D}^-_\alpha+\frac{{\rm i}}{2} ({\bar{\cal D}}^+_{\dot\alpha}{\bar
{\cal W}}){\bar{\cal D}}^{-{\dot\alpha}}- \frac{{\rm i}}{4}({\cal
D}^{+\a} {\cal D}^+_\a {\cal W}) D^{--}\non \\ &{}& +\frac{{\rm
i}}{8}[{\cal D}^{+\alpha},{\cal D}^-_\alpha] {\cal W} +
\frac{1}{2}\{{\bar {\cal W}},{\cal W} \}
\label{24}
\end{eqnarray}
Here index $\t$ denotes that corresponding superfields taken in
$\t$-frame \cite{GIKOS1} where covariant derivatives ${\cal
D}^i_{\alpha}$, $\bar{\cal D}^i_{\dot{\alpha}}$ are introduced to be
independent of harmonic coordinates. The propagators look like
(\ref{28}) just in this frame (see details in \cite{bko}).

Our aim consists in calculations of three- and four-loop
contributions to non-holomorphic effective potential.
We consider the case of the
gauge
group $SU(n)$ spontaneously broken down to its maximal Abelian
subgroup.
The corresponding background strength ${\cal W}$ is a diagonal matrix
of the form
\bea
\label{diag}
{\cal W}=diag({\cal W}_1,{\cal W}_2\ldots
{\cal W}_n);\ \sum_{i=1}^n {\cal W}_i=0.
\eea
Since non-holomorphic effective
potential depends only on background superfield strengths but not on
their derivatives we omit everywhere all terms including derivatives
of
${\cal W}$, $\bar{\cal W}$ in (\ref{28},\ref{24}).  Hence the
operator
${\stackrel{\frown}{\Box}}$ in propagators (\ref{28}) looks
like
\begin{eqnarray}
\label{24a}
{\stackrel{\frown}{\Box}}{}&=&{\cal D}^m{\cal D}_m+{\cal W}\bar {\cal
W},
\end{eqnarray}
or, in the manifest form,
\bea
\label{sun}
{\stackrel{\frown}{\Box}}{}&=&\left(\begin{array}{ccc}
{\cal D}^m{\cal D}_m+{\cal W}_1\bar{\cal W}_1&0&\ldots\\
0&\ldots&\ldots\\
0&\ldots&{\cal D}^m{\cal D}_m+{\cal W}_n\bar{\cal W}_n
\end{array}\right).
\eea
As a result we face a problem of calculating three- and four-loop
supergraphs in the theory with constant background superfield
strengths
and operator ${\stackrel{\frown}{\Box}}$ given by (\ref{diag})
and (\ref{24a}) respectively.

First of all, let us note that arbitrary $L$-loop supergraph provides
non-zero contribution to
non-holomorphic effective potential if and only if number
of $D$-factors contained in it is equal to $8L$ or greater since
contracting of any loop to a point by the rule\\
$\delta^{8}(\theta_1-\theta_2){(D^+(u_1))}^4{(D^+(u_2))}^4
\delta^{8}(\theta_1-\theta_2)={(u_1^+u_2^+)}^4\delta^{8}(\theta_1-\theta_2)$
requires 8 $D$-factors. Then, an arbitrary supergraph with
$P_v$ propagators of $N=2$ gauge superfield, $P_m$ propagators of
matter hypermultiplet, $P_c$ propagators of ghosts, $V_m$ vertices
containing interaction with matter and $V_c$ ones including
interaction with ghosts contains the following number of $D$-factors:
\bea
N_D=4P_v+8P_m+8P_c-4V_m-4V_c
\eea
because of structure of propagators given in (\ref{28}) with the
operator ${\stackrel{\frown}{\Box}}$ has the form (\ref{24a})
and vertices corresponding to actions (\ref{16},\ref{gh})
(recall that transformation of vertex
of the form $\int d\zeta^{(-4)}$ to an integral over whole superspace
by the rule $\int d\zeta^{(-4)}{(D^+)}^4=\int d^{12}z$ requires
four $D$-factors). Then, it is easy to see that number of propagators
of ghosts and matter is equal to number of vertices including
interaction with ghosts and matter respectively since each pure
ghost (matter) loop contains equal number of vertices and
propagators,
i.e. $P_m=V_m$, $P_c=V_c$. As a result, number of $D$-factors in
arbitrary supergraph is equal to $N_D=4P$ where $P$ is a full number
of propagators in corresponding supergraph. Hence any $L$-loop
supergraph can contribute to non-holomorphic effective potential if
and
only if $P\geq 2L$. For example, two-loop supergraph should contain 4
and more propagators, and since number of propagators in two-loop
supergraphs is no more than three we see that there
is no two-loop contribution to ${\cal H}({\cal W},\bar{\cal W})$
in accordance with conclusion of \cite{bko} where namely such an
analysis was used to prove absence of two-loop non-holomorphic
contribution.  However a situation beyond
two loops is much more complicated. In three- and four-loop
supergraphs number of propagators
should be no less than 6 or 8 respectively.
Such supergraphs actually exist and are studied bellow.

Let us consider three-loop supergraphs corresponding to theory
defined
by actions (\ref{16},\ref{gh}). They are given by Figs. 1a, 1b.
Here as usual wavy line is used for gauge propagator, solid line is for
matter propagator, dashed line is for ghost propagator. The numbers
near any supergraph at Fig. 1a will be explained later.

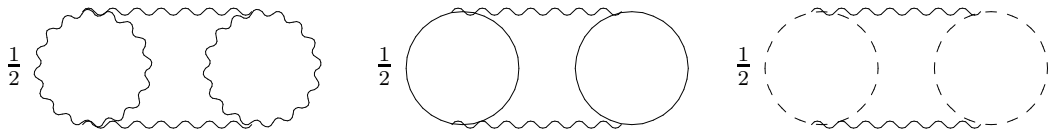
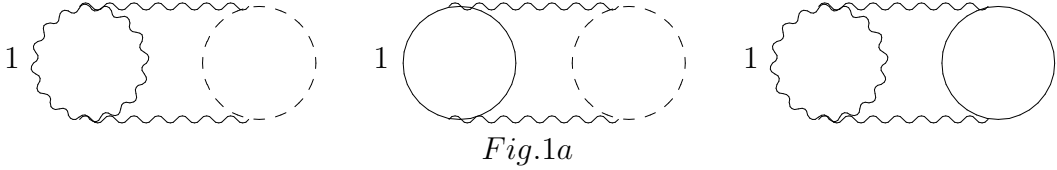
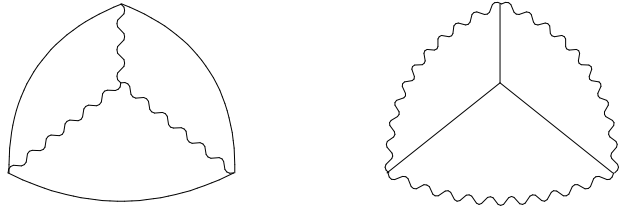
\Lengthunit=1.5cm
\hspace*{5mm}\GRAPH(hsize=3){
\wavelin(1.5,0)\mov(0,-1){\wavelin(1.5,0)}
\mov(0,-.5){\wavecirc(1)}\mov(1.5,-.5){\wavecirc(1)}\ind(-7,-5){\frac{1}{2}}
\mov(3,0){\GRAPH(hsize=3){
\wavelin(1.5,0)\mov(0,-1){\wavelin(1.5,0)}\ind(-6,-5){\frac{1}{2}}
\mov(0,-.5){\Circle(1)}\mov(1.5,-.5){\Circle(1)}
\mov(3,0){\GRAPH(hsize=3){
\wavelin(1.5,0)\mov(0,-1){\wavelin(1.5,0)}\ind(-6,-5){\frac{1}{2}}
\mov(0,-.5){\dashcirc(1)}\mov(1.5,-.5){\dashcirc(1)}
}}
}}
}

\vspace{4mm}

\hspace*{5mm}\GRAPH(hsize=3){
\wavelin(1.5,0)\mov(0,-1){\wavelin(1.5,0)}\ind(-6,-5){1}
\mov(0,-.5){\wavecirc(1)}\mov(1.5,-.5){\dashcirc(1)}
\mov(3,0){\GRAPH(hsize=3){
\wavelin(1.5,0)\mov(0,-1){\wavelin(1.5,0)}\ind(-6,-5){1}
\mov(0,-.5){\Circle(1)}\mov(1.5,-.5){\dashcirc(1)}\ind(6,-13){Fig.1a}
\mov(3,0){\GRAPH(hsize=3){
\wavelin(1.5,0)\mov(0,-1){\wavelin(1.5,0)}\ind(-6,-5){1}
\mov(0,-.5){\wavecirc(1)}\mov(1.5,-.5){\Circle(1)}
}}
}}
}

\vspace{4mm}

\hspace*{4.5cm}
\GRAPH(hsize=3){
\arcto(-1,-1.5)[-0.7]\arcto(1,-1.5)[0.7]
\mov(-1.1,-1.5){\arcto(2,0)[-0.5]}
\mov(-.1,-.7){\wavelin(0,.7)\wavelin(-1,-.8)\wavelin(1,-.8)}
\mov(3,0){
\GRAPH(hsize=3){
\wavearcto(-1,-1.5)[-0.7]\wavearcto(1,-1.5)[0.7]
\mov(-1.1,-1.5){\wavearcto(2,0)[-0.5]}\ind(-17,-21){Fig.1b}
\mov(-.2,-.7){\lin(0,.7)\lin(-1,-.8)\lin(1,-.8)}
}}
}

\vspace{4mm}

We are going to show that total contribution of the supergraphs given
by Fig. 1a to non-holomorphic effective potential vanishes due to
$N=4$ supersymmetry.  To clarify a mechanism providing manifestation of
$N=4$ supersymmetry in the supergraphs formulated in terms of $N=2$
superfields we introduce a notion of $N=4$ superpartner supergraphs.
Three supergraphs are called $N=4$ superpartners if they have the
following structure. One of the supergraphs contains the gauge loop
given by Fig.  2a and some system of the propagators associated with
this loop.  Another supergraph contains the matter loop given by Fig.
2c instead of gauge one and the same system of the propagators as in
first case.  Third supergraph contains the ghost loop given by Fig. 2b
instead of gauge one and the same system of the propagators as in first
case.  The examples of systems of superpartner supergraphs are given
also by Figs.  3a -- 3c, Figs. 4a -- 4c, Fig. 5.  Appearance of such a
set of supergraphs turned out to be typical for higher loop
contributions.  The simplest set of $N=4$ superpartner supergraphs
arising at one-loop order is given by Fig.  2a -- Fig. 2c.  We show
that sum of three these supergraphs for background dependent
superpropagators is equal to zero in the case of constant background
superfield strengths.

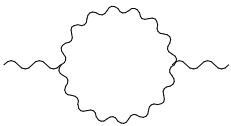
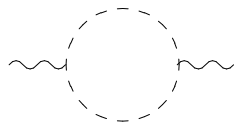
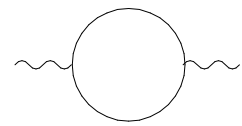
\vspace*{3mm}

\mov(-1.1,0){
\mov(3.0,0){\wavecirc(1)}
\mov(2.4,0){\wavelin(-0.5,0)}\ind(28,-8){Fig.2a}
\mov(3.3,0){\wavelin(0.5,0)}
\mov(3.0,0){
\mov(3.0,0){\dashcirc(1)}
\mov(2.4,0){\wavelin(-0.5,0)}\ind(28,-8){Fig.2b}
\mov(3.3,0){\wavelin(0.5,0)}
\mov(3.0,0){
\mov(3.0,0){\Circle(1)}
\mov(2.4,0){\wavelin(-0.5,0)}\ind(28,-8){Fig.2c}
\mov(3.3,0){\wavelin(0.5,0)}
}}}

\vspace*{3mm}

\noindent Structure of the
supergraphs containing $N=4$ superpartners in the case when
propagators do not depend on background superfields
 was studied in details by GIKOS [18]. However we prove that
the same result takes place when we consider
non-holomorphic effective potential using the background field
dependent propagators (7,10).

To evaluate contributions from supergraphs given by Figs. 2a -- 2c
one reminds that in $\tau$-frame (where the propagators look like
(\ref{28})) covariant harmonic derivatives $\nabla^{++},\nabla^{--}$
coincide with standard harmonic derivatives \cite{GIKOS1}. Then, the
commutation relations $[\nabla^{++},{\cal D}^{+}_A]=0$ with ${\cal
D}^{+}_A$ be either vector or spinor covariant derivative take
place in any frame \cite{bko}. At the same time we pay attention to
the
fact that
$[{\cal D}_{\gamma}^+,{\cal D}_{\alpha\dot{\alpha}}]=
i\epsilon_{\gamma\alpha}\bar{{\cal D}}^+_{\dot{\alpha}}\bar{\cal W}$.
Therefore one can put $[{\cal
D}_{\gamma}^+,{\stackrel{\frown}{\Box}}{}]=0$ in the sector of
non-holomorphic effective potential.

Let us consider the supergraphs given by Figs.2a -- 2c in more
details.  The contribution from the supergraph given at Fig.2a is
equal
to (see (\ref{28}))
\bea
I_a&=&\int d^8\theta_1 d^8\theta_2 \int
du_1du_2du_3dw_1dw_2dw_3 \frac{1}{\stackrel{\frown}{\Box}}
\delta^8_{12}{({\cal D}^+_2)}^4{({\cal D}^+_3)}^4\delta^8_{12}
V^{++}(1)V^{++}(2)
\times\nonumber\\&\times&
\frac{1}{\stackrel{\frown}{\Box}}
\frac{1}{(u^+_1u^+_2)(u^+_2u^+_3)(u^+_1u^+_3)(w^+_1w^+_2)(w^+_2w^+_3)(w^+_1w^+_3)}
\times\nonumber\\&\times&
\delta^{(-2,2)}(u_2,w_2)\delta^{(-2,2)}(u_3,w_3)
\eea
We take into account
that ${\cal D}^+$ commutes with $\stackrel{\frown}{\Box}$
and use the relation\\ $\delta^8_{12}{({\cal D}^+_2)}^4{({\cal
D}^+_3)}^4\delta^8_{12}= {(u^+_2u^+_3)}^4\delta^8_{12}$.
Integration over $w_2$ and $w_3$ leads to
\bea
I_a&=&\int d^8\theta\int
du_1du_2du_3dw_1\frac{1}{\stackrel{\frown}{\Box}^2}
V^{++}(1)V^{++}(2)
\frac{{(u^+_2u^+_3)}^2}
{(u^+_1u^+_2)(u^+_1u^+_3)(w^+_1u^+_2)(w^+_1u^+_3)}
\eea
This expression coincides with the result obtained in
\cite{GIKOS3}, the only difference consists in presence of
$\stackrel{\frown}{\Box}$ instead of $\Box$. However since
superfield strength ${\cal W}$ does not depend on harmonic
coordinates
we can carry out the trick which was used in \cite{GIKOS3}.
We express ${(u^+_2u^+_3)}^2$ as
$D^{++}_2D^{++}_3[(u^-_2u^-_3)(u^+_2u^+_3)]$ and integrate by
parts to transfer
$D^{++}_2D^{++}_3$ to other terms of $I_a$.
After that the $I_a$
takes the form
\bea
I_a&=&\int d^8\theta\int du_1dw_1\frac{1}{\stackrel{\frown}{\Box}^2}
V^{++}(1)V^{++}(2)
\frac{u^+_1w^-_1}
{u^+_1w^+_1}
\eea

Analogous consideration allows to show that contributions from
supergraphs given by Fig.2b, Fig.2c are respectively equal to
\bea
I_b&=&-2\int d^8\theta\int
du_1dw_1\frac{1}{\stackrel{\frown}{\Box}^2}
V^{++}(1)V^{++}(2)\frac{(u^+_1w^-_1)(u^-_1w^+_1)}{{(u^+_1w^+_1)}^2}
\eea
and
\bea
I_c&=&-2\int d^8\theta\int
du_1dw_1\frac{1}{\stackrel{\frown}{\Box}^2}
V^{++}(1)V^{++}(2)\frac{1}{{(u^+_1w^+_1)}^2}
\eea
Here $V^{++}(1)$, $V^{++}(2)$ are external gauge lines (which are
contracted to some systems of propagators or vertices in
multiloop
supergraphs containing graphs given at Fig.2a -- Fig.2c as
subdiagrams). These contributions are analogous to corresponding
expressions given in
\cite{GIKOS3} with the only difference in presence of
$\stackrel{\frown}{\Box}$ instead of $\Box$.
It is evident that
$I_1+I_2+I_3=0$ in accordance with \cite{GIKOS3}, i.e. these
supergraphs cancel each other in sector of constant
background fields. This effect provides vanishing of
sum of supergraphs given by Fig. 3a -- Fig. 3c, Fig. 4a -- Fig. 4c
and Fig. 5.

We note that vanishing of this sum is caused by $N=4$ supersymmetry
and does
not depend on structure of (Abelian) gauge group. Therefore such a
situation
is common for any $SU(n)$ gauge group broken down to its maximal
Abelian subgroup independently of value of $n$.

Let us return back to the supergraphs given by Fig. 1a. Each
supergraph has some combinatoric factor. The straightforward
calculations
show that they are proportional to each other with the coefficients
1/2 or 1 which are written near the corresponding supergraphs.  We call
these coefficients the relative factors.

The supergraphs given by Fig. 1a can be equivalently regroupped
into the sets of $N=4$ superpartners as shown on Figs. 3a - 3c where the
relative
factors are also written near the supergraphs. All supergraphs on
Fig. 3a
are $N=4$ superpartners as well as the supergraphs on Figs. 3b, 3c
respectivelly. As a result one gets the three sets of the $N=4$
superpartner
supergraphs. Each such a set can be studied by the same methos as the
supergraphs on Figs. 2a - 2c. It leads to conclusion that sum of the
contributions of the supergraphs in every set is equal to zero.

\Lengthunit=1.5cm
\hspace*{5mm}\GRAPH(hsize=3){
\ind(-6,-5){\frac{1}{2}}\wavelin(1.5,0)\mov(0,-1){\wavelin(1.5,0)}
\mov(0,-.5){\wavecirc(1)}\mov(1.5,-.5){\wavecirc(1)}
\mov(3,0){\GRAPH(hsize=3){
\ind(-6,-5){\frac{1}{2}}\wavelin(1.5,0)\mov(0,-1){\wavelin(1.5,0)}
\mov(0,-.5){\wavecirc(1)}\mov(1.5,-.5){\Circle(1)}\ind(7,-13){Fig.3a}
\mov(3,0){\GRAPH(hsize=3){
\ind(-6,-5){\frac{1}{2}}\wavelin(1.5,0)\mov(0,-1){\wavelin(1.5,0)}
\mov(0,-.5){\wavecirc(1)}\mov(1.5,-.5){\dashcirc(1)}
}}
}}
}

\vspace*{2mm}

\Lengthunit=1.5cm
\hspace*{5mm}\GRAPH(hsize=3){
\ind(-6,-5){\frac{1}{2}}\wavelin(1.5,0)\mov(0,-1){\wavelin(1.5,0)}
\mov(0,-.5){\dashcirc(1)}\mov(1.5,-.5){\wavecirc(1)}
\mov(3,0){\GRAPH(hsize=3){
\ind(-6,-5){\frac{1}{2}}\wavelin(1.5,0)\mov(0,-1){\wavelin(1.5,0)}
\mov(0,-.5){\dashcirc(1)}\mov(1.5,-.5){\Circle(1)}\ind(7,-13){Fig.3b}
\mov(3,0){\GRAPH(hsize=3){
\ind(-6,-5){\frac{1}{2}}\wavelin(1.5,0)\mov(0,-1){\wavelin(1.5,0)}
\mov(0,-.5){\dashcirc(1)}\mov(1.5,-.5){\dashcirc(1)}
}}
}}
}

\vspace*{2mm}

\Lengthunit=1.5cm
\hspace*{5mm}\GRAPH(hsize=3){
\ind(-6,-5){\frac{1}{2}}\wavelin(1.5,0)\mov(0,-1){\wavelin(1.5,0)}
\mov(0,-.5){\Circle(1)}\mov(1.5,-.5){\wavecirc(1)}
\mov(3,0){\GRAPH(hsize=3){
\ind(-6,-5){\frac{1}{2}}\wavelin(1.5,0)\mov(0,-1){\wavelin(1.5,0)}
\mov(0,-.5){\dashcirc(1)}\mov(1.5,-.5){\Circle(1)}\ind(7,-13){Fig.3c}
\mov(3,0){\GRAPH(hsize=3){
\ind(-6,-5){\frac{1}{2}}\wavelin(1.5,0)\mov(0,-1){\wavelin(1.5,0)}
\mov(0,-.5){\Circle(1)}\mov(1.5,-.5){\Circle(1)}
}}
}}
}

\vspace*{2mm}

Now we turn to remaining three-loop supergraphs including six
propagators which are given by Fig. 1b.  There are also two pairs of
supergraphs analogous to this pair, they include ghost and gauge
propagators instead of matter propagators.  After $D$-algebra
transformations contributions of both these supergraphs and their
analogs in which matter loop is replaced by gauge and ghost loop
respectively turn to be proportional to the following integral
over internal momenta:
\bea
\label{l3}
J&=&tr\int
d^8\theta\int\frac{d^4k_1 d^4k_2 d^4k_3}{{(2\pi)}^{12}}
\frac{1}{(k_1^2+{\cal W}\bar{\cal W})(k_2^2+{\cal W}\bar{\cal W})
(k_3^2+{\cal W}\bar{\cal W})}
\times\nonumber\\&\times&
\frac{1}{((k_1+k_2)^2+{\cal W}\bar{\cal W})
((k_1+k_3)^2+{\cal W}\bar{\cal W})((k_2+k_3)^2+{\cal W}\bar{\cal W})}
\eea
Here $tr$ is a matrix trace. We use the expression for $N=2$
background superfield strength in the form (\ref{diag}),
the operator $\stackrel{\frown}{\Box}$ in the form
(\ref{24a}).
The integral in the expression (\ref{l3}) is
formally logarithmically divergent (although supergraphs without
matter legs and legs including derivatives of gauge strengths have
superficial degree of divergence equal to zero \cite{bko}, the
corresponding divergences vanish due to supersymmetry).
Therefore we
carry out dimensional regularization via changing integration over
$d^4k_i$ by integration over $d^{4+\epsilon}k_i$ (with $i=1,2,3$).
Straightforward calculation of $J$ leads to the result
\bea
\label{j}
J=tr\frac{1}{{(16\pi^2)}^3}\int
d^8\theta(\frac{2}{\epsilon}+\log(\frac{{\cal W}\bar{\cal
W}}{\mu^2}))
\eea
However pole part vanishes due to known
properties of integration over anticommuting variables (see
f.e. \cite{BK0}).  Since ${\cal W},\bar{\cal W}$ are the diagonal
matrices they commute with each other, and we can use identity $\int
d^8\theta\log(\frac{{\cal W}\bar{\cal W}}{\mu^2})=\int
d^8\theta(\log(\frac{\cal W}{\mu})+\log(\frac{\bar{\cal W}}{\mu}))$.
This expression vanishes due to chirality of ${\cal W}$.
Therefore the supergraphs given by Fig. 1b
give zero contribution. The same situation takes place for analogous
supergraphs where the matter propagators are replaced by gauge
and ghost ones.  Hence we conclude that three-loop contribution to
non-holomorphic effective potential is equal to zero.

Now let us consider four-loop supergraphs. It turns to be that the
situation we observed at three-loop order takes place also at
four-loop
order, i.e. each supergraph either has $N=4$ superpartners sum
together with which it is equal to zero or gives the contribution
proportional to $\int d^{12}z\log(\frac{{\cal W}\bar{\cal
W}}{\mu^2})=0$.

First of all, at four-loop order we get two systems of supergraphs
given by Figs. 4.1 -- 4.6 and Fig.5
which can be separated into sets of $N=4$ superpartners.
The straightforward calculations show that the combinatoric factors of
different supergraphs at Figs. 4.1-4.6 are proportional to each other
with the relative factors 1 or 1/2.  These factors are written near of each
supergraph.

\vspace*{4mm}
\hspace*{5mm}\GRAPH(hsize=3){
\halfcirc(1)[l]\halfwavecirc(1)[r]\mov(0,.5){\lin(0,-1)}\ind(6,-8){Fig.4.1}
\mov(.86,0){\halfwavecirc(1)[l]\halfcirc(1)[r]\mov(0,.5){\lin(0,-1)}}
\ind(-9,0){\frac{1}{2}}
\mov(3,0){
\GRAPH(hsize=3){
\halfdashcirc(1)[l]\halfwavecirc(1)[r]\mov(0,.5){\dashlin(0,-1)}
\ind(6,-8){Fig.4.2}
\mov(.8,0){\halfwavecirc(1)[l]\halfdashcirc(1)[r]\mov(0,.5){\dashlin(0,-1)}}
\ind(-9,0){\frac{1}{2}}
\mov(3,0){
\GRAPH(hsize=3){
\halfwavecirc(1)[l]\halfwavecirc(1)[r]\mov(0,.5){\wavelin(0,-1)}
\ind(6,-8){Fig.4.3}
\mov(.8,0){\halfwavecirc(1)[l]\halfwavecirc(1)[r]\mov(0,.5){\wavelin(0,-1)}}
\ind(-9,0){\frac{1}{2}}
}}
}}
}

\vspace*{2mm}
\hspace*{5mm}\GRAPH(hsize=3){
\halfcirc(1)[l]\halfwavecirc(1)[r]\mov(0,.5){\lin(0,-1)}
\ind(6,-8){Fig.4.4}
\ind(-9,0){1}
\mov(.73,0){\halfwavecirc(1)[l]\halfdashcirc(1)[r]\mov(0,.5){\dashlin(0,-1)}}
\mov(3,0){
\GRAPH(hsize=3){
\halfdashcirc(1)[l]\halfwavecirc(1)[r]\mov(0,.5){\dashlin(0,-1)}
\ind(6,-8){Fig.4.5}
\ind(-9,0){1}
\mov(.73,0){\halfwavecirc(1)[l]\halfwavecirc(1)[r]\mov(0,.5){\wavelin(0,-1)}}
\mov(3,0){
\GRAPH(hsize=3){
\halfcirc(1)[l]\halfwavecirc(1)[r]\mov(0,.5){\lin(0,-1)}
\ind(6,-8){Fig.4.6}
\ind(-9,0){1}
\mov(.73,0){\halfwavecirc(1)[l]\halfdashcirc(1)[r]\mov(0,.5){\dashlin(0,-1)}}
}}
}}
}

\vspace*{4mm}

\noindent The scheme of separation
of supergraphs given by Fig. 4.1 -- Fig. 4.6 into sets of $N=4$
superpartners is given by Fig. 4a --
Fig. 4c where the relative factor is also manifestly shown near
corresponding supergraph.
We again get three sets of the $N=4$ superpartner supergraphs as in
three-loop case (see Figs. 3a - 3c).  Each line on the Figs. 4a -- 4c
contains the superpartner supergraphs.

\vspace*{4mm}
\hspace*{5mm}\GRAPH(hsize=3){
\halfcirc(1)[l]\halfwavecirc(1)[r]\mov(0,.5){\lin(0,-1)}
\ind(-9,0){\frac{1}{2}}
\mov(.8,0){\halfwavecirc(1)[l]\halfcirc(1)[r]\mov(0,.5){\lin(0,-1)}}
\mov(3,0){
\GRAPH(hsize=3){
\halfcirc(1)[l]\halfwavecirc(1)[r]\mov(0,.5){\lin(0,-1)}
\ind(-9,0){\frac{1}{2}}
\mov(.8,0){\halfwavecirc(1)[l]\halfdashcirc(1)[r]\mov(0,.5){\dashlin(0,-1)}}
\ind(4,-8){Fig. 4a}
\mov(3,0){
\GRAPH(hsize=3){
\halfcirc(1)[l]\halfwavecirc(1)[r]\mov(0,.5){\lin(0,-1)}
\ind(-9,0){\frac{1}{2}}
\mov(.8,0){\halfwavecirc(1)[l]\halfwavecirc(1)[r]\mov(0,.5){\wavelin(0,-1)}}
}}
}}
}

\newpage
\vspace*{4mm}
\hspace*{5mm}\GRAPH(hsize=3){
\halfdashcirc(1)[l]\halfwavecirc(1)[r]\mov(0,.5){\dashlin(0,-1)}
\ind(-9,0){\frac{1}{2}}
\mov(.8,0){\halfwavecirc(1)[l]\halfwavecirc(1)[r]\mov(0,.5){\wavelin(0,-1)}}
\mov(3,0){
\GRAPH(hsize=3){
\halfdashcirc(1)[l]\halfwavecirc(1)[r]\mov(0,.5){\dashlin(0,-1)}
\ind(-9,0){\frac{1}{2}}
\mov(.8,0){\halfwavecirc(1)[l]\halfdashcirc(1)[r]\mov(0,.5){\dashlin(0,-1)}}
\ind(4,-8){Fig. 4b}
\mov(3,0){
\GRAPH(hsize=3){
\halfdashcirc(1)[l]\halfwavecirc(1)[r]\mov(0,.5){\dashlin(0,-1)}
\ind(-9,0){\frac{1}{2}}
\mov(.8,0){\halfwavecirc(1)[l]\halfcirc(1)[r]\mov(0,.5){\lin(0,-1)}}
}}
}}
}


\vspace*{4mm}
\hspace*{5mm}\GRAPH(hsize=3){
\halfwavecirc(1)[l]\halfwavecirc(1)[r]\mov(0,.5){\wavelin(0,-1)}
\ind(-9,0){\frac{1}{2}}
\mov(.8,0){\halfwavecirc(1)[l]\halfcirc(1)[r]\mov(0,.5){\lin(0,-1)}}
\mov(3,0){
\GRAPH(hsize=3){
\halfwavecirc(1)[l]\halfwavecirc(1)[r]\mov(0,.5){\wavelin(0,-1)}
\ind(-9,0){\frac{1}{2}}
\mov(.8,0){\halfwavecirc(1)[l]\halfdashcirc(1)[r]\mov(0,.5){\dashlin(0,-1)}}
\ind(4,-8){Fig. 4c}
\mov(3,0){
\GRAPH(hsize=3){
\halfwavecirc(1)[l]\halfwavecirc(1)[r]\mov(0,.5){\wavelin(0,-1)}
\ind(-9,0){\frac{1}{2}}
\mov(.8,0){\halfwavecirc(1)[l]\halfwavecirc(1)[r]\mov(0,.5){\wavelin(0,-1)}}
}}
}}
}

\vspace*{4mm}

\noindent It is evident that the total contribution of the
supergraphs
given by Figs. 4a -- 4c is equal to that one of the supergraphs given
by Figs. 4.1 -- 4.6.
Since the supergraphs within
every set are the superpartners their total contribution vanishes.
It is proved by the same method as three loop order.


Another system of $N=4$ superpartner supergraphs at four-loop
order is given by Fig.5.

\vspace*{4mm}
\hspace*{10mm}
\GRAPH(hsize=3){
\halfwavecirc(2)[u]\mov(-1,0){\wavelin(2,0)}
\mov(-.96,-.4){\wavelin(1.87,.2)\wavelin(1.67,-.2)}
\mov(-.15,0){\halfwavecirc(2)[d]}
\mov(3,0){\GRAPH(hsize=3){
\halfdashcirc(2)[u]\mov(-1,0){\dashlin(2,0)}
\mov(-.96,-.4){\wavelin(1.87,.2)\wavelin(1.67,-.2)}
\ind(0,-14){Fig.5}
\mov(-.25,0){\halfwavecirc(2)[d]}
\mov(3.3,0){\GRAPH(hsize=3){
\halfcirc(2)[u]\mov(-1,0){\lin(2,0)}
\mov(-.96,-.4){\wavelin(1.87,.2)\wavelin(1.67,-.2)}
\mov(-.15,0){\halfwavecirc(2)[d]}
}}
}}
}

\vspace*{2mm}

\noindent Since these supergraphs are $N=4$ superpartners, it is easy
to see that sum of contributions of these supergraphs is equal to
zero.

The remaining four-loop supergraphs with eight propagators are given
by
Figs. 6a -- 6c.

\vspace*{4mm}

\hspace*{10mm}
\GRAPH(hsize=3){
\Circle(2)\mov(1,0){\wavelin(-2,0)}\mov(0,1){\wavelin(0,-2)}
\ind(0,-14){Fig.6a}
\mov(3,0){\GRAPH(hsize=3){
\halfwavecirc(2)[d]\wavelin(-1,0)\lin(1,0)\lin(0,1)\mov(-1,0){\wavelin(1,.5)}
\mov(-1.1,0){\wavearcto(1,1)[.7]}\mov(.9,0){\arcto(-1,1)[-.7]}
\ind(0,-14){Fig.6b}
}
\mov(.4,0){\GRAPH(hsize=3){
\mov(-0.9,0){\wavearcto(1,-1)[-.7]}\mov(1.1,0){\wavearcto(-1,-1)[.7]}
\wavelin(-1,0)\lin(1,0)\lin(0,1)\mov(-1,0){\wavelin(1,-1)}
\mov(-1.1,0){\wavearcto(1,1)[.7]}\mov(0.9,0){\arcto(-1,1)[-.7]}
\ind(0,-14){Fig.6c}}
}
}
}

\vspace*{4mm}

\noindent The contributions of these supergraphs after
$D$-algebra transformations are proportional
to the following integrals over internal momenta respectively:
\bea
J_{6a}&=&tr\int d^8\theta\int\frac{d^4k_1 d^4k_2 d^4k_3
d^4k_4}{{(2\pi)}^{16}}
\frac{1}{(k_1^2+{\cal W}\bar{\cal W})(k_2^2+W\bar{W})(k_3^2+{\cal
W}\bar{\cal W})}
\times\nonumber\\&\times&
\frac{1}{((k_1+k_2)^2+{\cal W}\bar{\cal W})((k_1+k_2+k_3)^2+{\cal
W}\bar{\cal W})
((k_1+k_4)^2+{\cal W}\bar{\cal W})}
\times\nonumber\\&\times&
\frac{1}{(k^2_4+{\cal W}\bar{\cal W})((k_2+k_3-k_4)^2+{\cal
W}\bar{\cal W})}\nonumber\\
J_{6b}&=&tr\int d^8\theta\int\frac{d^4k_1 d^4k_2 d^4k_3
d^4k_4}{{(2\pi)}^{16}}
\frac{1}{(k_1^2+{\cal W}\bar{\cal W})(k_2^2+{\cal W}\bar{\cal W})
(k_3^2+{\cal W}\bar{\cal W})}
\times\nonumber\\&\times&
\frac{1}{((k_1+k_2)^2+{\cal W}\bar{\cal W})((k_1+k_3)^2+{\cal
W}\bar{\cal W})
((k_3+k_4)^2+{\cal W}\bar{\cal W})}
\times\nonumber\\&\times&
\frac{1}{(k^2_4+W\bar{W})((k_1+k_3+k_4)^2+{\cal W}\bar{\cal
W})}\nonumber\\
J_{6c}&=&tr\int d^8\theta\int\frac{d^4k_1 d^4k_2 d^4k_3
d^4k_4}{{(2\pi)}^{16}}
\frac{1}{(k_1^2+{\cal W}\bar{\cal W})(k_2^2+{\cal W}\bar{\cal W})
(k_3^2+{\cal W}\bar{\cal W})}
\times\nonumber\\&\times&
\frac{1}{((k_1+k_2)^2+{\cal W}\bar{\cal W})((k_2+k_3)^2+{\cal
W}\bar{\cal W})
((k_2+k_3-k_4)^2+{\cal W}\bar{\cal W})}
\times\nonumber\\&\times&
\frac{1}{(k^2_4+{\cal W}\bar{\cal W})((k_1+k_2+k_4)^2+{\cal
W}\bar{\cal W})}
\eea
After dimensional regularization and integration the $J_{6a}$,
$J_{6b}$, $J_{6c}$ turn to be equal to
\bea
J_{6a}=J_{6b}=J_{6c}=tr\frac{1}{{(16\pi^2)}^4}\int
d^8\theta(\frac{2}{\epsilon}+\log(\frac{{\cal W}\bar{\cal
W}}{\mu^2}))
\eea
This expression is completely analogous to (\ref{j}). As a result we
get the same conclusion. Each of
supergraphs given by Figs. 6a -- 6c is proportional to
$\int d^{12}z\log(\frac{{\cal W}\bar{\cal W}}{\mu^2})$ where
${\cal W}$ is diagonal matrix of the form (\ref{diag}) and
$\bar{\cal W}$ is its conjugate.  The same situation takes place
for the
analogous supergraphs where the matter superpropagators are replaced
by
gauge and ghost ones. Hence their contributions also vanish.  By the
way, we convinced that terms of the form $\frac{{\cal W}_a-{\cal
W}_b}{{\cal W}_c-{\cal W}_d}$ supposed in [8,15] do not arise in
non-holomorphic effective potential at least at three and four loops.

We investigated all four-loop supergraphs including
eight internal lines.  Namely for this
number of superpropagators all $D$-factors are used for contracting
loops to points in $\theta$-space. The four-loop supergraphs
with nine superpropagators are also present but after $D$-algebra
transformations in such supergraphs the extra factor ${({\cal
D}^+)}^4$
remains. It can act only on background superfield strengths.
Therefore four-loop supergraphs with nine propagators cannot
contribute
to non-holomorphic effective potential.

To conclude, we have considered three- and four-loop supergarphs
contributing to non-holomorphic effective potential and proved that
only two situations
are possible for these supergraphs: (i) either contribution of this
supergraph is proportional to $\int d^{12} z \log
(\frac{{\cal W}\bar{\cal W}}{\mu^2})$, and such a structure vanishes
due to
properties of integral in superspace (ii) or such a supergraph has
$N=4$ superpartner supergraphs, and sum of contribution from three
$N=4$ superpartner supergraphs is equal to zero because of $N=4$
supersymmetry. This result is common for any unitary gauge group
broken down to its maximally symmetric torus since vanishing of sum
of
superpartner supergraphs is caused only by $N=4$ supersymmetry, not
by
structure of gauge group.

We found that the mechanism of vanishing of corrections
to non-holomorphic effective potential at three-loop order
essentially
differs from that one at two loops.  Absence of two-loop
contributions
is stipulated only by $N=2$ supersymmetry.  $N=4$ supersymmetry
begins
to work efficiently at three loops and higher
and manifests itself by means of $N=4$ superpartner
supergraphs.
However we proved that
the situation at four loops is completely analogous to one in the
previous order. The mechanism of vanishing the three- and four-loop
contributions to non-holomorphic effective potential looks like very
generic and one can expect that it works at any loop.

Detailed study of the structure of above three- and four- loop
supergraphs and explicit results of their calculations will
be published elsewhere.

{\bf Acknowledgements.}
The authors are glad to acknowledge S.M. Kuzenko for valuable
discissions and remarks and A.A. Tseytlin for interest to the work
and remarks.
The work was supported in part by the RFBR grant \symbol{242}
99-02-16617,
by the DFG-RFBR grant \symbol{242} 99-02-04022, by the INTAS grant
No 96-0308 and by GRACENAS grant \symbol{242} 97-6.2-34.
I.L.B. is grateful to the NATO research grant PST.CLG 974965,
to the FAPESP grant
and to Institute of Physics, University of S\~{a}o Paulo for
hospitality.

\end{document}